\documentclass[twocolumn,aps,floatfix]{revtex4}
\usepackage{amssymb} \usepackage{graphicx} \usepackage{amsmath}
\usepackage[T1]{fontenc} \usepackage{pstricks} \usepackage{subfigure}

\begin{document}
\title{Density fluctuations in a quasi-one-dimensional Bose gas as observed in free expansion}

\author{Krzysztof Gawryluk,$^{1,2}$ Mariusz Gajda,$^{3,4}$ and
  Miros{\l}aw Brewczyk$\,^{1,4}$}

\affiliation{\mbox{$^1$Wydzia{\l} Fizyki, Uniwersytet w
    Bia{\l}ymstoku,  ul. K. Cio{\l}kowskiego 1L, 15-245 Bia{\l}ystok,
    Poland}  \\ \mbox{$^2$Centre for Quantum Technologies, National
    University of Singapore, 3 Science Drive 2, Singapore 117543,
    Singapore }  \\ \mbox{$^3$Institute of Physics PAN, Al. Lotnik\'ow
    32/46, 02-668 Warsaw, Poland}  \\ \mbox{$^4$Center for Theoretical
    Physics PAN, Al. Lotnik\'ow 32/46, 02-668 Warsaw, Poland}  }

\date{\today}

\begin{abstract}
We study, within a framework of the classical fields approximation, the
density correlations of a weakly interacting expanding Bose gas for
the whole range of temperatures across the Bose-Einstein condensation
threshold. We focus  on elongated quasi-one-dimensional systems where there is a
huge discrepancy between the existing theory and experimental results
(A. Perrin et al., Nature Phys. 8, 195 (2012)). We find that the
density correlation function is not reduced for temperatures below the
critical one as it is predicted for the ideal gas or for a weakly
interacting system within the Bogoliubov approximation. This behavior of the
density correlations agrees with the above mentioned experiment with 
the elongated system. Although the system  was much larger then studied here  
we believe that the behavior of the density correlation function 
found there is quite generic. Our theoretical studies 
indicate also large density fluctuations in the trap in the quasicondensate regime
where only phase fluctuations were expected. We argue that the enhanced density
fluctuations can originate in the presence of interactions in the
system, or more precisely in  the existence of spontaneous dark
solitons in the elongated gas at thermal equilibrium.

\end{abstract}

\maketitle

Correlations are the essence of any many body systems. In particular,
the quantum features of the system are manifested by some unusual
correlations. The first order coherence is the basic criterion used
as the definition of a Bose-Einstein condensate. And indeed, very soon after
observation of trapped atomic condensates the first order coherence of such
systems, manifesting itself in the ability to produce the
interference fringes in  a two-slit experiment, had been proven
\cite{Ketterle} experimentally.  Higher order correlations leading
to atom bunching  were established from collisions and three-body losses 
\cite{Burt}. Although the Glauber coherence theory
introduced to characterize correlations of quantum electromagnetic
field is well established now, the issue of coherence 
of a matter field is still under intensive investigation. 

A Bose-Einstein condensate is a matter wave analogue of a coherent light. A very important question is how far this analogy can be pursuit. The first difference is that  atoms exist in a Fock states only: the states which are the eigenstates of the particle number operator. As a consequence, the coherent state of a matter field, understood as the exact analogue of the coherent electromagnetic field, does not exist.  On the other hand, the atomic field can exhibit a higher order coherence too. The coherence ought to be understood here as the ability to produce the interference patterns not only in a single-particle detection schemes, but also in a simultaneous detection of larger number of atoms. Although the phase of a number state can be arbitrary, the Fock states can interfere  in a single realization of the system \cite{Juha} when no averaging over global phase is performed. Therefore higher order coherence of atomic condensates, as can be observed in a single realization of the system, is an interesting issue and in fact this is the property of the system which characterizes a genuine Bose-Einstein condensate  \cite{Gajda}. The correlation functions of a Bose-Einstein condensate -- a coherent atomic wave -- can be contrasted with the correlations of fermionic atoms and/or thermal nondegenerate clouds. 

The famous experiment by R. Hanbury Brown and R.Q. Twiss \cite{HBT56} exploring  a new type of interferometer to measure the angular diameter of radio stars initiated  a modern theory of coherence. The Hanbury Brown and Twiss effect is rooted in  properties of the density-density correlations or equivalently, the second order correlation function.  The bunching effect for thermal photons originating from many particle interference enhancing probability of many photons detection can be contrasted to the absence of such an effect for coherent light when only a single mode is populated.  The atom counting experiments with metastable bosonic or fermionic helium atoms, exploring thermal density-density correlations, show bunching of nondegenerate bosons and  antibunching effects for fermions \cite{Westbrook}. Similar results were obtained in experiments with  nondegenerate gas at optical lattices \cite{Bloch} and atomic lasers \cite{Esslinger}. Direct measurement of the second and the third order correlation function of Bose-Einstein condensates is an important proof of their high coherence \cite{Truscott}.

The correlation functions of the atomic field,  contrary to the electromagnetic waves,  depend not only on the temperature, but also on the interactions and the dimensionality of the system. The density and phase of a three-dimensional condensate do not fluctuate. The situation is different in lower dimensions. In a two-dimensional Bose gas the Berezinskii-Kosterlitz-Thouless transition is manifested by a characteristic behavior of correlation functions \cite{Hadzibabic06}. In this paper we concentrate on a quasi-one-dimensional system, i.e. the situation when both: the chemical potential $\mu$ as well as the thermal energy $k_BT$ are smaller than a transverse confinement energy $\mu, k_B T \lesssim \hbar \omega_r$, where $\omega_r$ is the transverse trap frequency. 

One-dimensional systems are different than their 2D or 3D analogues. First of all there is no phase transition from the thermal to the condensate phase both in a trap and a uniform system. There exists however a degeneracy temperature $T_d=\hbar^2 n^2/2m$ ($n$ is a 1D density) below which the quantum effects become important in the uniform gas at the thermodynamic limit. Similarly, for the ideal 1D gas confined in the harmonic trap of frequency $\omega$, the characteristic temperature of quantum degeneracy is equal to $T_c=(\hbar \omega/k_B) N/\log(2N)$. If, in addition to a harmonic confinement, the gas is weakly interacting, then there exist a second characteristic temperature $T_{\phi}=15(\hbar\omega_z)^2N/32\mu < T_c$. As shown in \cite{Petrov} for temperatures range in between the two $T_{\phi}< T < T_c$, the system is in a quasicondensate regime where density fluctuations are suppressed but phase fluctuations are large. Only below $T_{\phi}$ the system enters a fully coherent region with suppressed density and phase fluctuations. The phase fluctuations have been observed in expansion when they are transformed into density fluctuation \cite{Sengstock}. The statement of suppressed density fluctuations remains a common believe as there is no direct experimental evidence supporting or disclosing the  above mentioned theoretical prediction based on the Bogoliubov approach.

For a very elongated quasi-one-dimensional system the local density approximation is justified. Then both: the trapped and the uniform systems should exhibit similar properties. The uniform system interacting via the zero-range pseudo-potential of the strength $g$ is formally exactly soluble via the Bethe ansatz.  In the limit of strong interaction, i.e. if $\gamma=mg/\hbar^2n \gg 1$ the system reaches the Tonks-Girardeau limit of impenetrable bosons i.e. the regime of fermionization. The system we investigate here is in the opposite regime of parameters, namely the regime where interactions are small.

One-dimensional Bose gases are special systems in a sense that, as
discovered many years ago by Lieb \cite{LiebII}, they exhibit two
families of elementary excitations. As already identified by Lieb
\cite{LiebII}, the main branch is related to phonons. It turns out
that dark solitons can be associated to the type II branch of the
elementary excitations. That has been demonstrated by analyzing the
dispersion relation for solitary waves \cite{DS0} and also by 
studies of quantum nature of dark solitons beyond the mean-field 
\cite{DS1}. Recently, we have shown that dark solitons are
spontaneously generated in quasi-one-dimensional Bose gases at
equilibrium \cite{Karpiuk12}. By analyzing the statistical
distributions of excitations within both the Lieb-Liniger model
\cite{LiebI} and the classical fields approximation \cite{CFA, review}
we proved that type II excitations are indeed quantum solitons
\cite{Karpiuk15}.

The spectrum of elementary excitations plays a crucial role for the correlation functions and coherence of the system. There is a reach literature devoted to both the first order \cite{Kadio,Bienias} and the second order correlations \cite{Kheruntsyan03,Kheruntsyan05,Sykes08,Deuar09} of the 1D or quasi-one-dimensional systems. A number of different approximate formulas for the second order correlation function in various regimes of the system parameters and temperatures are available. In particular in the region of fermionization the antibunching of  atoms is predicted. On contrary, at temperatures close to the degeneracy temperature and in the limit of weak interactions, the bunching of atoms is expected. The piece-wise valid formulas, although quite interesting, do not give a simple and clear picture of the density-density correlations in the entire range of temperatures across the `BEC-like transition' for a weakly interacting system where the Bogloliubov theory adopted to elongated system can be compared to  the predictions based on the Lieb-Liniger model. 

Observation of a spatial dependence of the second order correlation functions is difficult for several reasons. First of all observations in situ are limited by a finite resolution of detectors and, in addition, the  standard detection techniques involve a partial spatial averaging -- therefore a column density is monitored only.  The most common method of detection is destructive and the atomic cloud is observed after releasing atoms from the trap followed by the expansion.  All these effects smear out most of density fluctuations. Earlier studies of the second order correlation function in expansion considered  ballistic expansion  of noninteracting atomic cloud \cite{Gomes06} showing that the correlation function observed is related to the correlation in situ by some scaling only. In \cite{Imambekov09} the authors showed that density-density correlations of the interacting uniform gas manifest themselves in the expansion in a form of  density ripples. The spectrum of the ripples is related to density-density correlations. The recent experimental results  of \cite{Perrin12} for the two-point density correlation measured across the Bose-Einstein condensation temperature in elongated expanding system are still awaiting full understanding and theoretical description. The authors support the experimental results with the ideal gas theory and evidently one should account for interactions to get a deeper insight into measured density correlations.

In this paper we are going to demonstrate that the properties of the
density correlations of a Bose gas at temperatures below the critical
one, after the gas is released from the trap, indicate large density 
fluctuations in the quasicondensate regime and might be a signature of
the presence of thermal solitons in the system. As our calculations
show, for elongated Bose gases the density correlation function takes
large values below the threshold temperature, i.e. for $T< T_c$. 
These values are larger
than those calculated for a weakly-interacting Bose gas within the
Bogoliubov approximation and differ even more significantly from the
results obtained for the ideal Bose gas theory. This indicates that
not only phonons but also the second branch of elementary excitations
influences the properties of density correlations. Large values of the
density correlation function are therefore a signature of the presence
of dark solitons. This should not be very surprising as an every dark 
soliton is associated with a density dip. The recent experiment 
showing large values of the density
correlation function for temperatures below the critical one 
\cite{Perrin12} (see Fig. 3c) support our finding.

Below we are going to describe our numerical approach to determine the 
density-density correlation function of the elongated interacting Bose
gas after expansion from the harmonic trap, taking into account a realistic 
detection procedure, in particular a column averaging and finite resolution of a CCD camera.

In second quantization formalism one introduces the field operator
$\hat{\Psi}({\bf{r}})$ $(\hat{\Psi}^\dagger({\bf{r}}))$ which
annihilates (creates) an atom at position $\bf{r}$. The second order
correlation function, $G^{(2)} (\bf{r},\bf{r'})$, is then defined as a
statistical average
\begin{equation}
G^{(2)} ({\bf{r},\bf{r'}}) = \langle  \hat{\Psi}^\dagger({\bf{r}})
\hat{\Psi}({\bf{r}})  \hat{\Psi}^\dagger({\bf{r'}})
\hat{\Psi}({\bf{r'}})   \rangle    \,.
\label{G2}
\end{equation}
It is convenient to use a normalized second order correlation function
\begin{equation}
g^{(2)} ({\bf{r},\bf{r'}}) = \frac{G^{(2)} ({\bf{r},\bf{r'}})}
{G^{(1)} ({\bf{r},\bf{r}})  G^{(1)} ({\bf{r'},\bf{r'}})  }    \,,
\label{g2}
\end{equation}
where $G^{(1)} ({\bf{r},\bf{r'}}) = \langle
\hat{\Psi}^\dagger({\bf{r}}) \hat{\Psi}({\bf{r'}}) \rangle$ is the
first order correlation function.

To calculate the second order correlation function we turn to the
approximate treatment of an interacting Bose gas, which is called the
classical fields method (CFA) \cite{CFA,review}. The idea of the  approach can be
considered as an extension of the original Bogoliubov idea
\cite{Bogoliubov}, i.e. the bosonic field operator
$\hat{\Psi}({\bf{r}})$ is replaced by the complex wave function
$\Psi({\bf{r}})$. This wave function corresponds to a state having large energy
related to the temperature and is composed of  macroscopically occupied
single particle modes only. To obtain the thermal equilibrium state of an
elongated weakly interacting Bose gas we generate an ensemble of
classical fields $\Psi({\bf{r}})$ corresponding to a given
temperature. An effective way of getting members of such a canonical
ensemble is to use the Monte Carlo algorithm \cite{Witkowska}.

Each classical field belonging to the canonical ensemble obeys the
following equation of motion \cite{review}:
\begin{eqnarray}
&&i\hbar \frac{\partial}{\partial t} {\Psi }({\rm {\bf r}},t) = \left[
    -\frac{\hbar^2}{2m} \nabla^2 + V_{trap}({\rm {\bf r}},t)
    \right] {\Psi }({\rm {\bf r}},t)   \nonumber  \\ &&+ g\, {\Psi
  }^*({\rm {\bf r}},t) {\Psi }({\rm {\bf r}},t) {\Psi }({\rm {\bf
      r}},t)   \,,
\label{CFequ}
\end{eqnarray}
where the coupling constant $g$ describes the contact interaction
between atoms. It looks like the usual Gross-Pitaevskii equation 
for the Bose-Einstein condensate at zero
temperature. However, here the complex function $\Psi({\bf{r}})$
carries information on both the condensed and non condensed atoms. The
condensate and the thermal cloud can be split via the coarse-graining
procedure \cite{review, Karpiuk10}. In short, the coarse-graining allows to separate
the mode characterized by the largest first order coherence length (or a coherence time)
from the remaining part of the classical field.

Different realizations of classical fields, being the members of the
canonical ensemble, represent the set of atomic clouds at the thermal
equilibrium. While evolving with the help of Eq. (\ref{CFequ}) the
classical field reveals the properties of an atomic cloud within a
single-shot experiment. We therefore pick up typically thousands of
members of the canonical ensemble and propagate them  without an
external trapping potential, i.e. $V_{trap}=0$, for a given time. We
then compute the autocorrelation function of the density integrated
along one of the radial directions $\tilde{n}_{2D}(x,z)=\int
|\Psi(x,y,z)|^2 dy$, where $z$ is the axial direction. However,
because of the finite resolution of the optical imaging system we
smooth the radially integrated density via the following
convolution:
\begin{eqnarray}
n_{2D}({\bf{r}}) = A \int \tilde{n}_{2D}({\bf{u}})
\exp{\left[-({\bf{r}-\bf{u}})^2/2\sigma^2 \right]} \, d^2u  \,,
\label{conv}
\end{eqnarray}
with the Gaussian function of the half width at half maximum (HWHM) of
about a few micrometers and $A=1/2\pi\sigma^2$.  The density
correlation function itself is obtained by averaging the
autocorrelation $\int n_{2D}({\bf{u}})  n_{2D}({\bf{u}+\bf{r}}) \,
d^2u$ over all performed realizations:
\begin{equation}
G^{(2)} (0,{\bf{r}}) = \langle  \int n_{2D}({\bf{u}})
n_{2D}({\bf{u}+\bf{r}}) \, d^2u  \,\rangle  \,.
\label{G2num}
\end{equation}
To get the normalized density correlation function we divide
(\ref{G2num}) by the autocorrelation function of the mean density \cite{Manz10}, integrated along the radial direction:
\begin{equation}
g^{(2)} (0,{\bf{r}}) = G^{(2)} (0,{\bf{r}}) \,\, / \int   \langle
n_{2D}({\bf{u}}) \rangle \, \langle n_{2D}({\bf{u}+\bf{r}}) \rangle \,
d^2u   \,.
\label{g2norm}
\end{equation}
Autocorrelation functions appearing in both the numerator and
denominator of the above definition (\ref{g2norm}) are calculated
using the fast Fourier transform.

\begin{figure}[thb] 
\includegraphics[width=6.3cm]{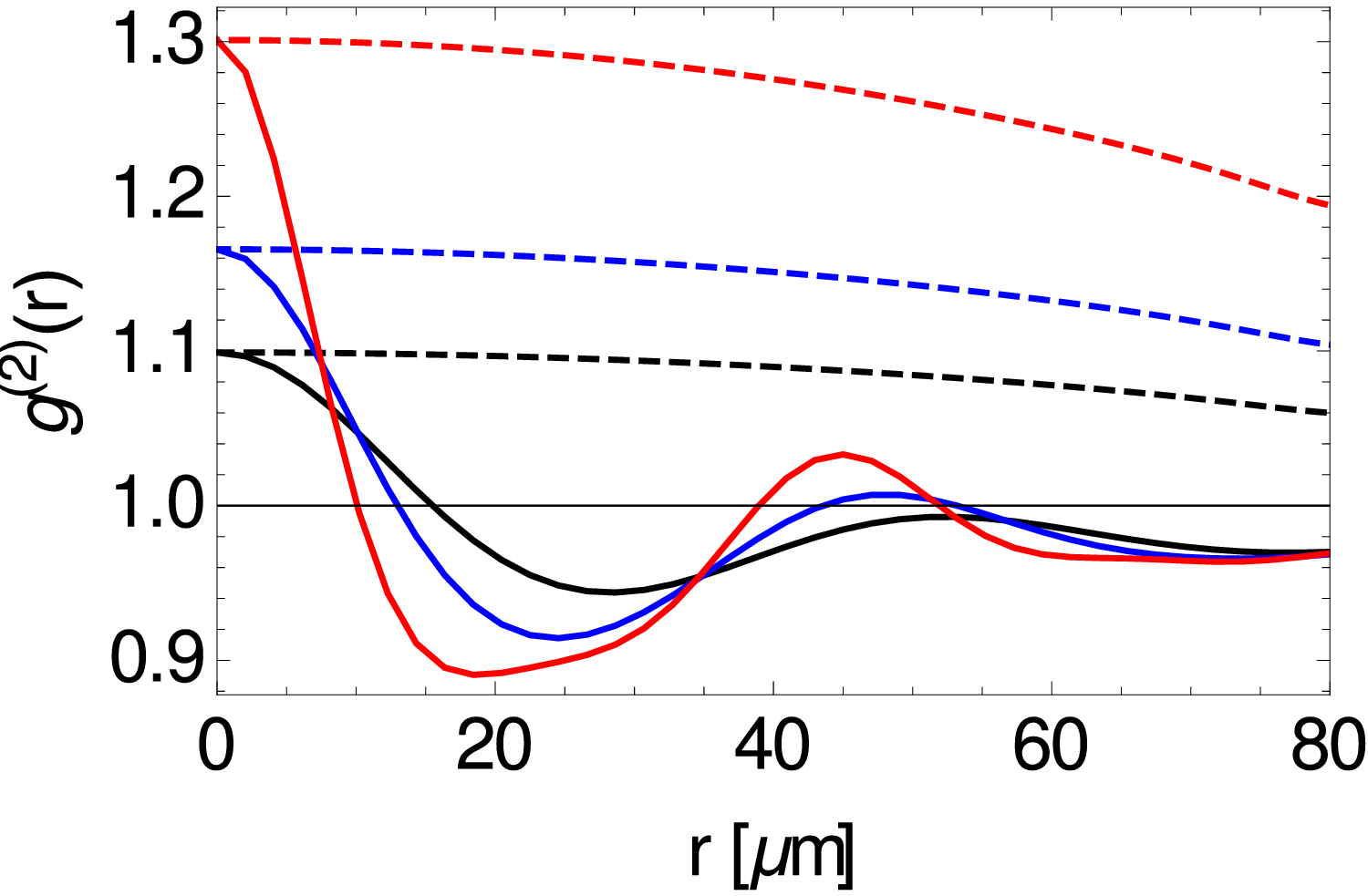}  \\  \vspace{0.4cm}
\includegraphics[width=4.1cm]{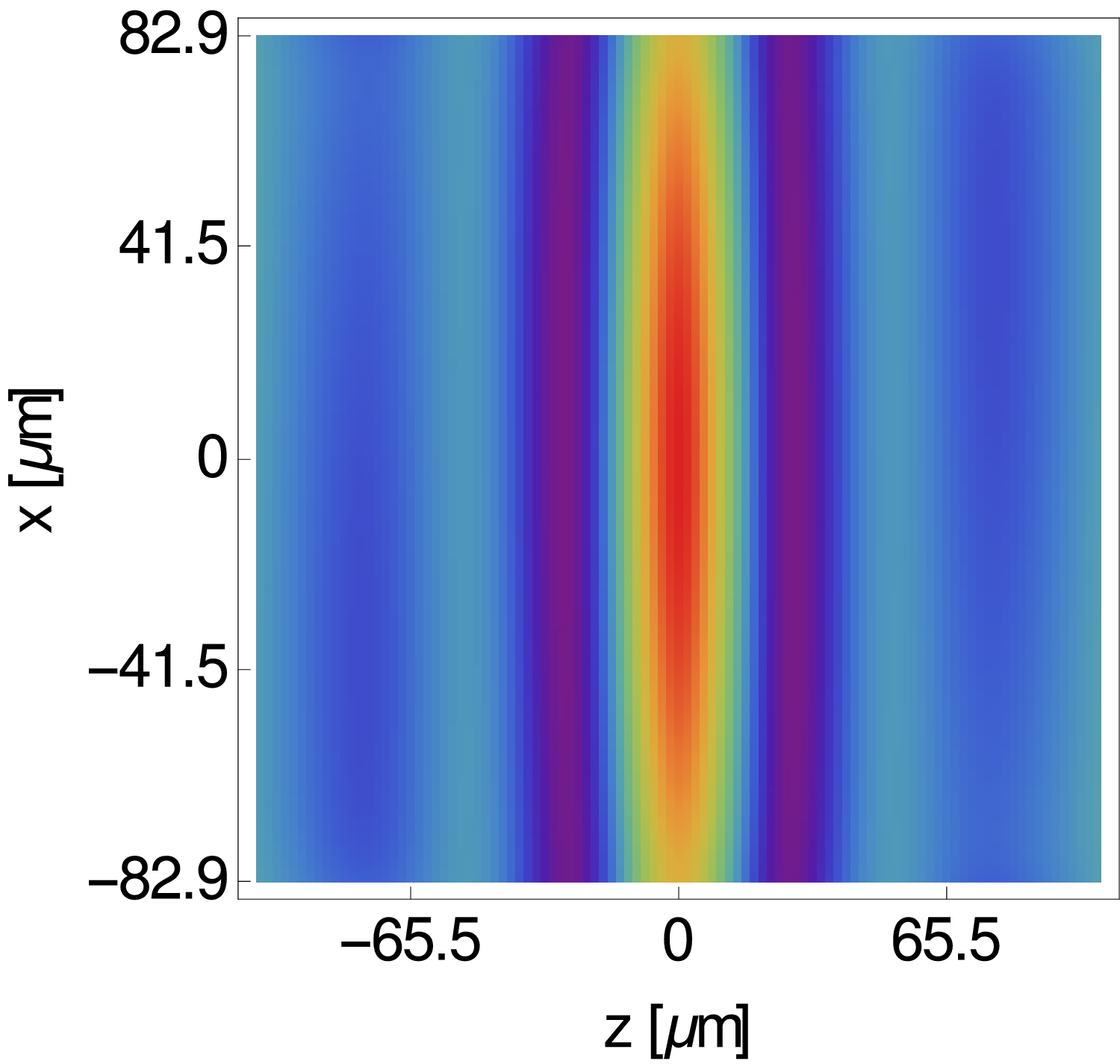}  \hfill
\includegraphics[width=4.1cm]{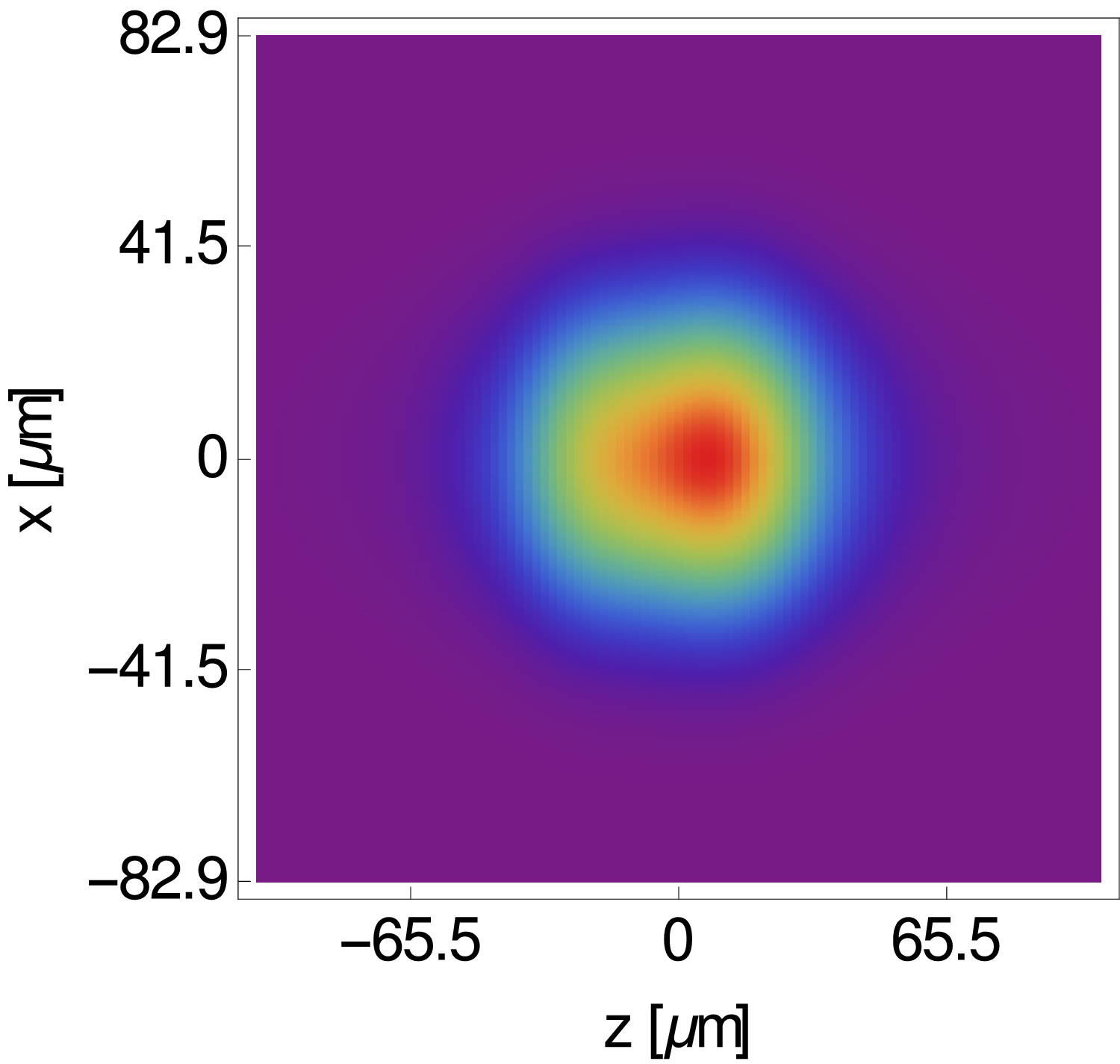} 
\caption{(color online). Axial (solid lines) and radial (dashed lines)
  cuts of the density correlation function $g^{(2)}(x,z)$ of the Bose
  gas after $46\,$ms of expansion (upper frame). The Bose gas
  consisting of $10^3$ rubidium atoms is initially confined in a
  harmonic trap with radial and axial frequencies $\omega_r=2\pi
  \times 113\,$Hz and $\omega_z=2\pi \times 1\,$Hz, respectively. The
  system is prepared in the thermal equilibrium at the temperature
  $T=3.15\,$nK. Three sets of curves are obtained by taking different
  values of the smoothing parameter: $\sigma=4\,\mu$m, $6\,\mu$m, and
  $8\,\mu$m (from top to bottom).  Lower panel: The density
  correlation function $g^{(2)}(x,z)$ for $\sigma=8\,\mu$m (left
  frame) and the two-dimensional density (i.e., density averaged along
  the direction of imaging) after expansion, first smoothed with the
  parameter $\sigma=8\,\mu$m and next averaged over all realizations
  (right frame).  }
\label{g2_275}
\end{figure}

Because we use the classical fields approximation, i.e. we
replace  the bosonic fields operators by the complex functions, the
shot-noise term does not appear in our approach. The origin of the
shot-noise term is in the commutation relations between the creation
and annihilation operators for the bosonic field. Obviously in the
classical fields approximation the bosonic commutation relations are
not accounted for. It might be questionable to use the CFA for
obtaining the density correlation function because the shot-noise term
appears explicitly in the expression for the second order correlation
function while expressed  with the help of the first order correlation
functions. The shot-noise term is of the order of the number of
particles in the system, $N$, whereas the other terms are proportional
to $N^2$. Hence, it can be neglected for large systems. Although, it
is detectable in the experiment of Ref. \cite{Perrin12}, a special
care is taken to exclude the atomic shot-noise peak from the
experimental data.

In Fig. \ref{g2_275} we show the density correlation function,
$g^{(2)} (x,z)$, for the temperature $T=3.15\,$nK after free expansion
over $46\,$ms (upper and lower-left frames). The Bose gas consisting
of $10^3$ rubidium atoms is initially confined in a
quasi-one-dimensional harmonic trap with the radial frequency
$\omega_r=2\pi \times 113\,$Hz and the axial one $\omega_z=2\pi \times
1\,$Hz. The aspect ratio is just as for the most elongated trap used
in the experiment of Ref. \cite{Perrin12}. Similarly to the
experiment, the theoretical system is also in the
quasi-one-dimensional limit, i.e. its chemical potential is smaller
than the radial excitation energy, $\hbar \omega_r$. Some other
parameters are different, in particular the number of atoms is $10$
times smaller. We use the system with smaller number of atoms confined
in a trap with lower trap frequencies to decrease its mean field
energy. In such a case the expansion of the gas after the trap is
removed is much slower and becomes feasible numerically. We believe,
however, that the  scaling does not alter significantly the main
physical processes responsible for behavior of the density correlation
function.

The gas is then released from the trap and expanded for $46\,$ms. For
that we solve the Eq. (\ref{CFequ}) with no trapping potential,
$V_{trap}=0$. The expansion is not rapid and after $46\,$ms the radial
size of the cloud remains at the order of tens micrometers (see
lower-right frame in Fig. \ref{g2_275}, where we show the density,
first smoothed with the parameter $\sigma=8\,\mu$m and then averaged
over all realizations). As opposed to the experiment of
Ref. \cite{Perrin12} in our calculations we image the whole radial
extension of the cloud. The upper frame in Fig. \ref{g2_275} presents the 
density correlation function cuts along the axial (solid curves) and radial
(dashed curves) directions. The bunching at short interparticle
distance ($g^{(2)}>1$) as well as the oscillations of correlations along 
axial direction with the values below one are clearly visible. Our results
correspond to the observations in the experiment of
Ref. \cite{Perrin12} (see also earlier experimental data of Ref. \cite{Manz10}). 
At large distances which coincide with the edge of the atomic cloud, 
the density correlation function reaches its limit equal to one.

Speaking more precisely, during the expansion, in the limit of large 
expansion time, we monitor the momentum
distribution, i.e., the Fourier transform of the classical field. We
find that it does not change in time almost at all  what means that the
interaction energy does not play any significant role during the
expansion in our case, i.e., for the system with $N=1000$ atoms. In practice
the classical field evolves freely and at any time
can be found with the help of the propagator of the free Schr\"odinger
equation. This technique is much faster than direct solving of the
Eq. (\ref{CFequ}) on the grid and, as we checked, gives the same results. In fact, this technique was used by us
since while calculating the density correlation function, the Eq. (\ref{g2norm}) 
has to be averaged over a few thousand of realizations.

\begin{figure}[thb] 
\includegraphics[width=6.3cm]{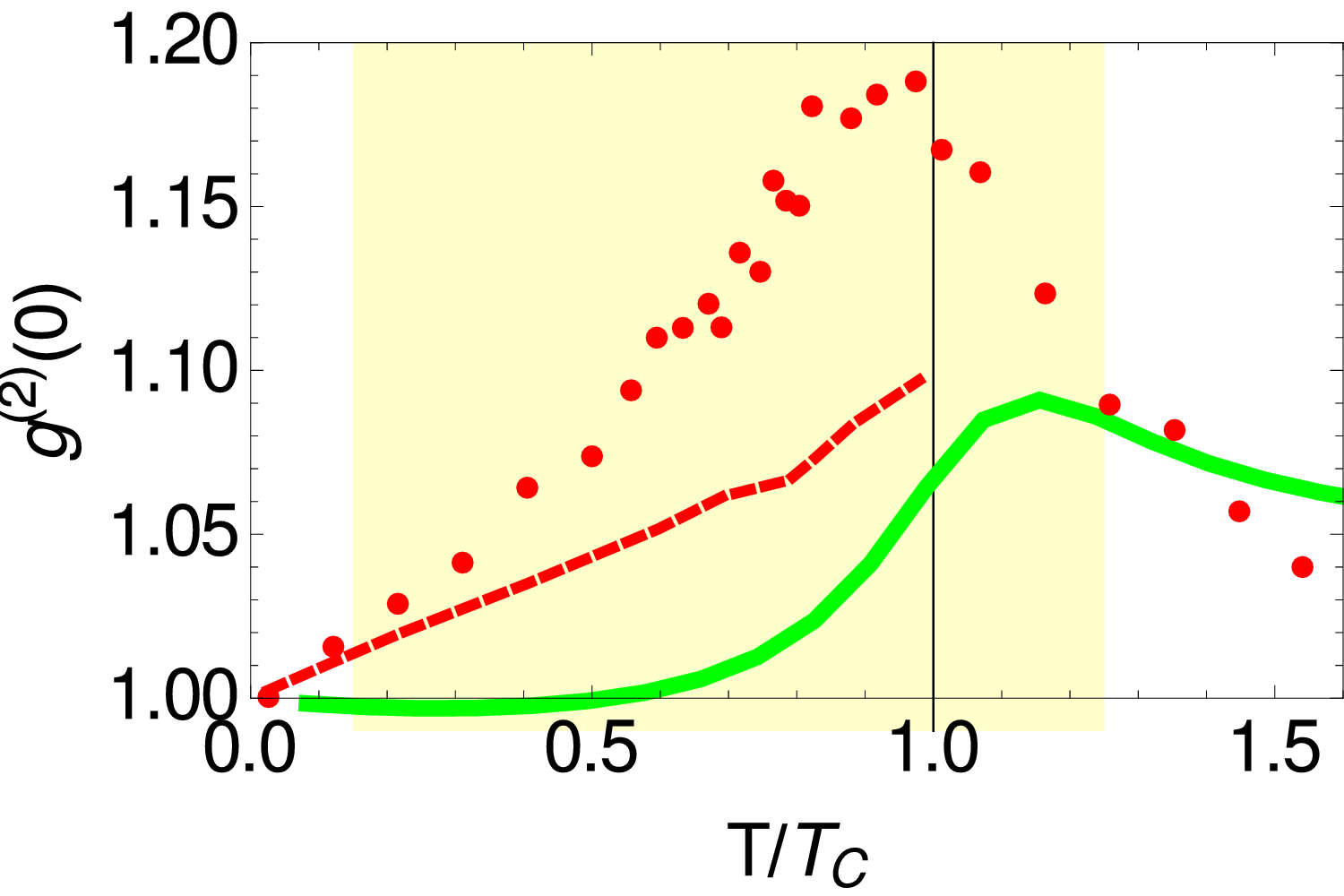} \\   \vspace{0.4cm}
\includegraphics[width=6.0cm]{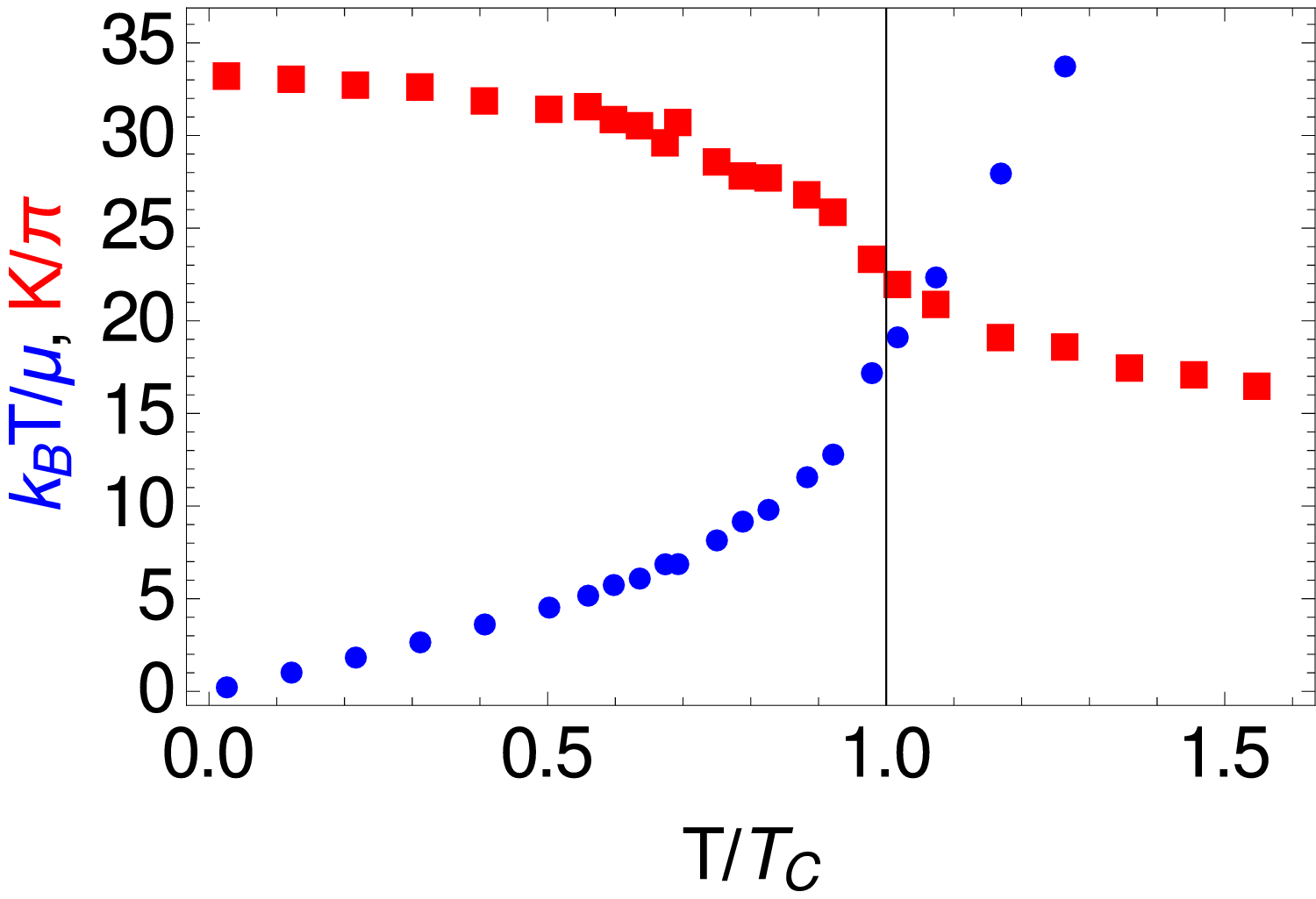}
\caption{(color online). Upper frame: Peak height of the density correlation
  function as calculated within the classical field approximation (red
  dots), the ideal Bose gas model as in Ref. \cite{Gomes06} (green
  solid line), and the Bogoliubov approximation extended to
  low-dimensional quasicondensates as in \cite{Imambekov09} (red
  dashed line). The critical temperature equals $T_c=6.3\,$nK and the
  smoothing parameter $\sigma=8\,\mu$m. The shaded area shows the
  range of temperatures when deep spontaneous dark solitons appear in
  the system. Note large values of the density correlation function
  below the threshold temperature calculated within the classical
  field approximation. Lower frame: $k_B T/ \mu$ (blue dots) and 
  $K/\pi$ (red squares) as functions of temperature to confront the validity
  of the model of Ref. \cite{Imambekov09} (red dashed line in the upper frame). }
\label{main}
\end{figure}

Next, we analyze the peak value of the second order correlation
function, $g^{(2)}(0,0)$, of expanding Bose gas for various
temperatures. There is a huge discrepancy between the experimentally
measured values and the results predicted by the ideal Bose gas theory
for elongated systems, see Ref. \cite{Perrin12}. For temperatures
below the critical one the peak height of $g^{(2)}(x,z)$ rapidly
decreases to unity for the ideal gas case whereas experiment proves
that the peak height does not change significantly within a large range of
temperatures. In Fig. \ref{main} we mark the results obtained within
the classical field approximation by red dots. The
temperature is given in units of the critical temperature for the
ideal Bose gas corrected due to finite number of atoms and the
quasi-one-dimensional character of the trapping potential. For the
ideal Bose gas in the thermodynamic limit the critical temperature
is $T_c=\hbar \bar{\omega} N^{1/3} / k_B \left[\zeta(3)\right]^{1/3}$,
where $\bar{\omega}=(\omega_r^2 \omega_z)^{1/3}$ and $\zeta()$ is the
Riemann zeta function \cite{PethickSmith}. For the parameters we use
in our calculations, $T_c=10.5\,$nK. After corrections mentioned above
the critical temperature is, however, shifted down and equals
$T_c=6.3\,$nK \cite{PethickSmith}. Since in the system we consider
the number of atoms is very small, the interactions do not influence
the critical temperature much.

Now we compare our results with the ones obtained within the
theoretical model described in Ref. \cite{Imambekov09}. This model
allows to calculate the two-point density correlation function of very
elongated ultracold Bose gas after its release from the trap. 
For a tight radial confinement the transverse atomic motion is essentially 
frozen and thus decouples from the motion along the long axis. Therefore,
the problem of calculating the density correlations is effectively reduced 
to 1D. The density correlation function turns out to be at any time analytically
related to the spectrum of ``density ripples'' which, on the other hand, can be 
obtained within the Bogoliubov approximation extended to low-dimensional
quasicondensates. The shape of the density correlation function depends on the 
expansion time, temperature, and the 1D density. The 1D density is calculated as the 
average of the radially integrated central densities of classical fields belonging to the 
canonical ensemble corresponding to the given temperature. The red dashed line in
Fig. \ref{main}, upper frame, is the output of such calculations. One has to remember, 
however, that the validity of Ref. \cite{Imambekov09}'s model is restricted 
to such temperatures that $k_B T/ \mu  \ll K/\pi$, where $K=\pi \hbar
n_{1D}/(m c)$ is the Luttinger liquid parameter with $c$ being the speed
of sound. In Fig. \ref{main}, lower frame, we show both quantities, 
$k_B T/ \mu$ and $K=\pi$, to visualize the above mentioned condition.
Certainly, the results of this model are questionable for temperatures about 
the critical one. On the other hand, for lower temperatures it is clear that 
the interactions strongly influence the density correlation function.

\begin{figure}[thb]
\includegraphics[width=4.1cm]{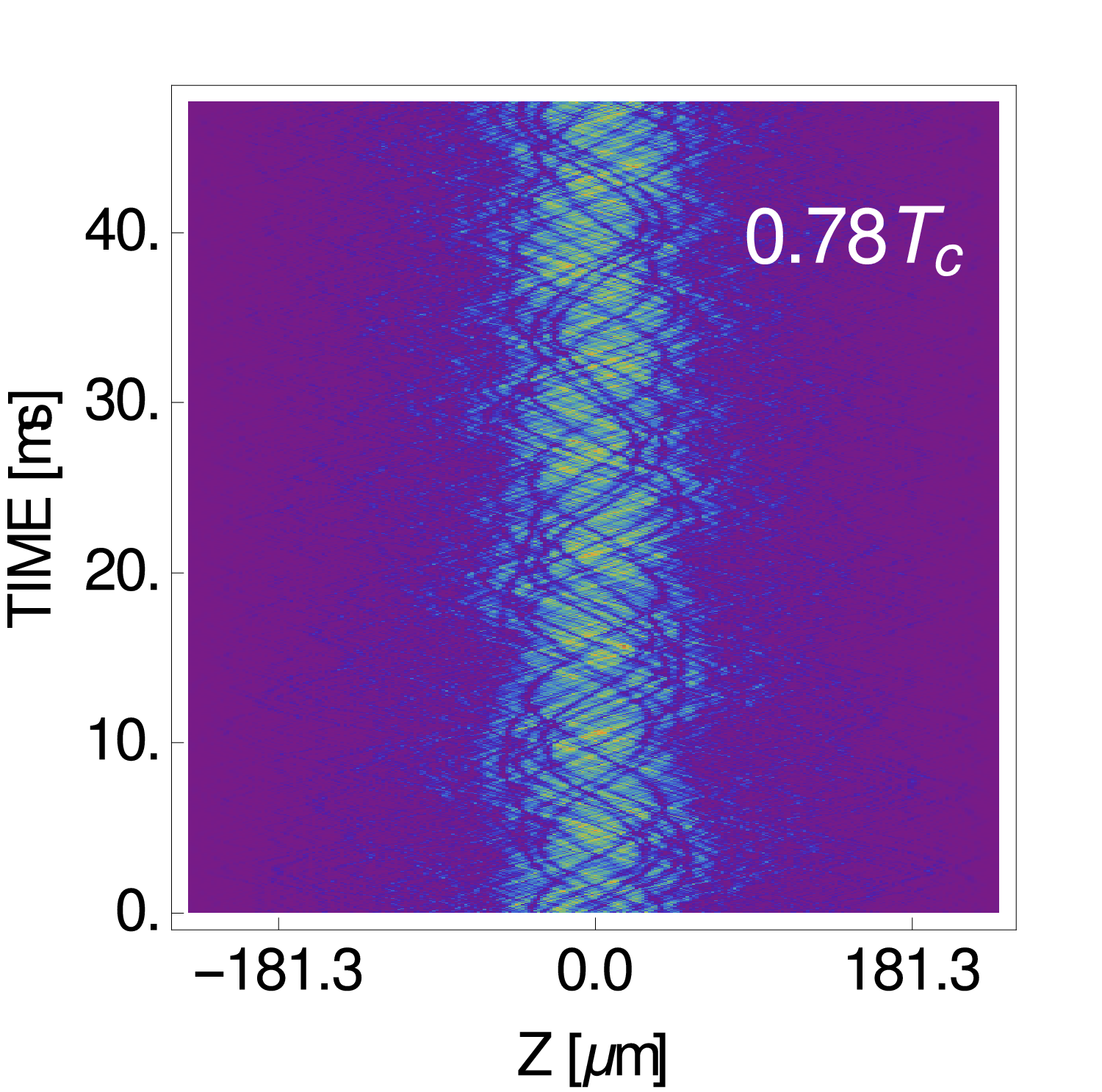}  \hfill
\includegraphics[width=4.1cm]{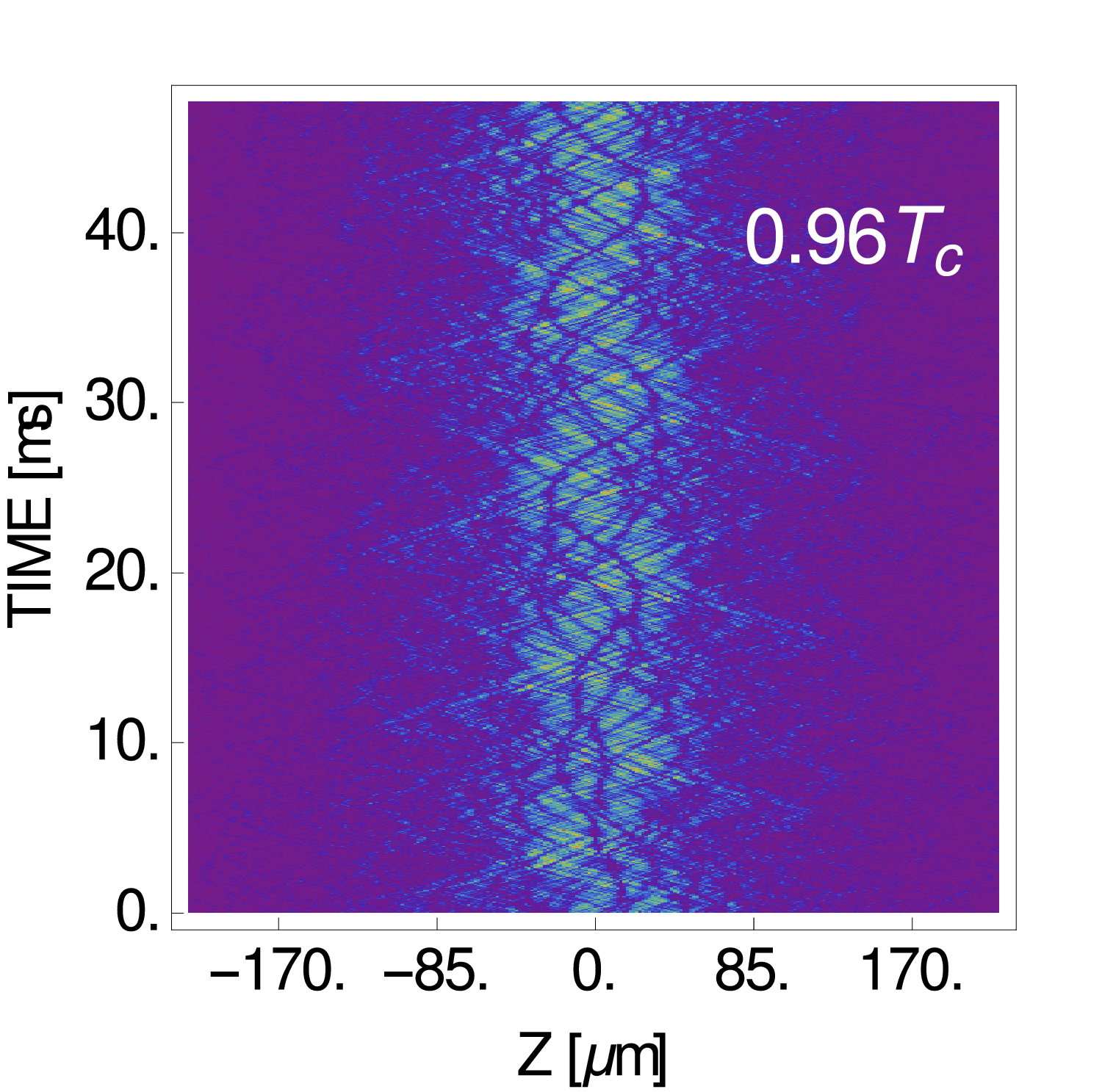}
\caption{Solitons for trapped atomic clouds at thermal equilibrium for
  the temperatures $T=4.94\,$nK and $T=6.09\,$nK. The panels show the
  time evolution of the density along the axial direction. Dark
  solitons are clearly visible as dark lines winding from the one edge 
  of atomic cloud to the other. }
\label{solitons}
\end{figure}

Fig. \ref{main} clearly demonstrates the differences between
theoretical predictions based on the ideal Bose gas model as well as
on the model of an interacting gas within the Bogoliubov approximation
and the ones obtained in the framework of the classical fields
approximation. Only CFA results qualitatively recover the experimental
data of Ref. \cite{Perrin12} showing the significant density
correlations below the critical temperature for the elongated Bose
gases. The reason is that CFA reproduces the correct spectrum of elementary excitations
including not only the Bogoliubov phonons but also the II type excitations -- the 
dark solitons. They do occur spontaneously in a
quasi-one-dimensional interacting Bose gas at equilibrium as it was
already reported in \cite{Karpiuk12}. The dark solitons do not affect the first
order correlation function, i.e. the coherence length as shown in
\cite{Karpiuk12}, however they affect the density fluctuations. Large
peak values of density correlations can be associated with the
existence of solitons, see Fig. \ref{solitons}. As it was already
shown many years ago by Lieb \cite{LiebI,LiebII} and later on
generalized to the finite temperature case by Yang and Yang
\cite{Yang}, a one-dimensional system exhibits two families of
excitations. The one branch belongs to phonons and is well described
by the Bogoliubov approximation. This kind of excitations are included
in the model of Ref. \cite{Imambekov09}. However, there is the second
branch of excitations which are dark solitons \cite{Karpiuk15} which
is neglected in the considerations in Ref. \cite{Imambekov09}. Hence,
the origin of the discrepancy between the two approaches is, in our
opinion, related to the presence of the second type of excitations
in  one-dimensional systems, i.e. to the dark solitons. They are
especially important in the regime of a quasicondensate where very
deep solitons are possible due to large thermal energy.

Above the critical temperature the peak density correlation should be equal
to two, indicating atom bunching. The CFA calculations, however, give
a value of the second order correlation function close to one rather then two. This is
related to the fact that the characteristic length over which the
particles bunching vanishes (i.e. $g^{(2)}(0)=1$) decreases with increasing temperature. 
Eventually, this length becomes smaller than
the spatial resolution of the imaging system and the bunching effect cannot be 
visible any more. Our calculations 
take into account the realistic resolution of detectors, therefore
they give a decrease of the correlations above the critical temperature as it is observed
in the experiment \cite{Perrin12}. In our case, the decrease of
$g^{(2)}(0)$ is slower since we have fewer number of atoms in the
system.

In summary, we have studied the density correlation function of
expanding Bose gas. The weakly interacting Bose gas at thermal equilibrium
is confined initially in a very elongated trap. The whole range of temperatures 
is considered, from temperatures as low as those of a pure condensate  through
the ones typical for quasicondensates up to temperatures much above the critical
one. Below the critical temperature the normalized density
correlation function does not fall rapidly to the value one as expected for the coherent 
system and predicted for the ideal Bose gas \cite{Gomes06}. 
Our finding sheds a new light onto the understanding of coherence of 
elongated or quasi-one-dimensional system. Contrary to what was 
expected, relatively large phase and density fluctuations signify
a lack of a full first and second order coherence and atom bunching
in a large range of temperatures below the degeneracy temperature.
Indeed, the
interactions are responsible for these features. However, the theory
based on the Bogoliubov approximation alone seems not to be able to
fix the problem. This is because in elongated systems besides the
phonon excitations described by the Bogoliubov approximation there
exist another excitations as predicted by the Lieb-Liniger
model. These are the dark solitons. Our approach captures both kinds of
excitations and therefore qualitatively recovers the experimental data
of Ref. \cite{Perrin12}.

\acknowledgments  We are grateful to A. Perrin and K. Rz\k{a}\.zewski
for helpful discussions.  The work was supported by the National
Science Center (Poland) Grants No. DEC-2011/01/B/ST2/05125 (K.G. and
M.B.) and DEC-2012/04/A/ST2/00090 (M.G.). The Centre for Quantum
Technologies is a Research Centre of Excellence funded by the Ministry
of Education and the National Research Foundation of Singapore.

\end{document}